\documentclass[twocolumn,prl,aps,showpacs,amsmath,amssymb]{revtex4}

\usepackage{graphicx}
\usepackage{dcolumn}
\usepackage{bm}

\renewcommand{\vec}[1]{{\bf#1}}

\begin{document}

\title{Quantum dissipative Rashba spin ratchets}

\author{Sergey Smirnov,$^1$ Dario Bercioux,$^{1,2}$ Milena Grifoni,$^1$ and Klaus Richter$^1$}
\affiliation{$^1$Institut f\"ur Theoretische Physik, Universit\"at Regensburg, D-93040 Regensburg, Germany\\
  $^2$Physikalisches Institut, Albert-Ludwigs-Universit\"at, D-79104 Freiburg, Germany}

\date{\today}

\begin{abstract}
We predict the possibility to generate a finite stationary spin current by applying an unbiased ac driving to a
quasi-one-dimensional asymmetric periodic structure with Rashba spin-orbit interaction and strong dissipation. We show
that under a finite coupling strength between the orbital degrees of freedom the electron dynamics at low temperatures
exhibits a pure spin ratchet behavior, {\it i.e.} a finite spin current and the absence of charge transport in spatially
asymmetric structures. It is also found that the equilibrium spin currents are not destroyed by the presence of strong
dissipation.
\end{abstract}


\maketitle

An opportunity to induce a net stationary particle current by unbiased external forces applied to a quantum dissipative
one-dimensional (1D) periodic structure is provided when the system does not possess a center of inversion in real space
\cite{Reimann}. Then the particle transport occurs due to the ratchet effect and the device works as a Brownian motor
\cite{Astumian}. In the deep quantum regime the charge ratchet effect can only be achieved when at least the two lowest
Bloch bands contribute to transport \cite{Grifoni}.

Recently a new research field of condensed matter physics, spintronics, has emerged. One of its central issues is how to
generate pure spin currents (SC) in paramagnetic systems due to only spin-orbit interactions and without applied magnetic
fields. Rashba spin-orbit interaction (RSOI) \cite{Rashba} represents one of the possible tools to reach this goal since
the spin-orbit coupling strength can be externally controlled by a gate voltage. One way to get pure SC is due to the
intrinsic spin Hall effect \cite{Murakami,Sinova} expected in a high-mobility two-dimensional semiconductor systems with
RSOI \cite{Wunderlich}. Such pure SC were experimentally detected through the reciprocal spin-Hall effect in Ref.
\cite{Valenzuela}. An alternative is to induce pure SC through absorption of polarized light \cite{Zhou}. The generation
of pure SC by coherent spin rectifiers \cite{Scheid} has been discussed only recently for a finite size setup with RSOI.
However, the presence of dissipation  has not been considered up to now.
\begin{figure}
\includegraphics[width=5.8 cm]{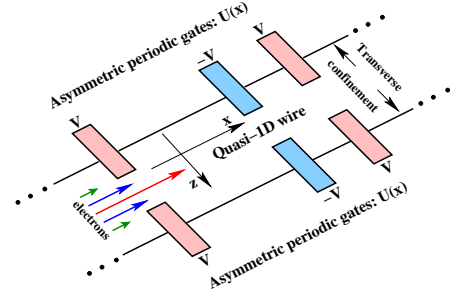}
\caption{\label{figure_1} (Color online) A schematic picture of the isolated asymmetric periodic quasi-1D structure
described by the Hamiltonian (\ref{isolated_hamiltonian}). In the center of the quasi-1D wire the periodic potential is
weaker and gets stronger closer to the edges. Thus the electron group velocity is higher in the central region and tails
off away from the center.}
\end{figure}

In this letter we address the challenging task of how to implement devices which can work both as Brownian charge and spin
motors. Here a natural and also principle question for spintronics arises: Is it possible to switch a device working as a
charge ratchet to a pure spin ratchet mode where the charge current (CC) is completely blocked? As mentioned above, when
in a dissipative system without RSOI transport is restricted to only one Bloch band, the charge ratchet mechanism does not
exist \cite{Grifoni}. Whether the same effect takes place in a dissipative system with RSOI is an open and non-trivial
question. In fact, the Rashba Hamiltonian is not invariant under reflection of a transport direction. Thus the Rashba
Hamiltonian itself already has a built-in spatial asymmetry which due to the spin-orbit coupling can be further mixed with
the periodic potential symmetry/asymmetry. The presence of dissipation additionally increases the complexity of the
problem because the influence of a dissipative environment on the orbital motion changes through RSOI the spin dynamics.

In this work  we focus on the moderate-to-strong dissipation case and address how to implement a device which under
influence of unbiased external ac-driving yields a finite stationary spin current and at the same time blocks the directed
stationary charge transport. To concretize our idea of a Brownian spin motor we consider a dissipative periodic system
with RSOI and show that the spin-orbit interaction alone is not enough to produce SC: The system must additionally lack
the spatial symmetry and its orbital degrees of freedom must be coupled.

The full Hamiltonian of our problem is $\hat{H}_{\text{full}}(t)=\hat{H}+\hat{H}_{\text{ext}}(t)+\hat{H}_\text{bath}$, where
$\hat{H}$ is the Hamiltonian of the isolated periodic system, $\hat{H}_{\text{ext}}(t)$ describes an external driving, and
$\hat{H}_\text{bath}$ is responsible for dissipative processes.

The isolated quasi-1D periodic system is formed in a two-dimensional electron gas (2DEG) with RSOI using a periodic
potential along the $x$-axis and a harmonic confinement along the $z$-axis:
\begin{equation}
\begin{split}
\hat{H}\!\!=\!\!\frac{\hbar^2\hat{\vec{k}}^2}{2m}\!+\!\frac{m\omega_0^2\hat{z}^2}{2}\!-\!
\frac{\hbar^2k_{\text{so}}}{m}\bigl(\hat{\sigma}_{x}\hat{k}_z-\hat{\sigma}_z\hat{k}_x\bigl)+U_\gamma(\hat{x},\hat{z}),
\end{split}
\label{isolated_hamiltonian}
\end{equation}
where $U_\gamma(\hat{x},\hat{z})=U(\hat{x})(1+\gamma\hat{z}^2/L^2)$, $\hat{\vec{k}}$ is related to
the momentum operator as $\hat{\vec{p}}=\hbar\hat{\vec{k}}$, $\omega_0$ is the harmonic confinement strength, $k_\text{so}$
the spin-orbit coupling strength, $U(\hat{x})$ the periodic potential with the period $L$, and $\gamma\geqslant 0$ the
orbit-orbit coupling strength. This isolated structure is sketched in Fig.~\ref{figure_1} as it could be realized by
appropriate gate evaporation techniques applied to 2DEGs formed in III-V compounds.

The periodic structure is subject to an external homogeneous time-dependent electric field,
$\vec{E}(t)\equiv E(t)\hat{e}_x$. It can be experimentally implemented using for example linearly polarized light. This
yields $\hat{H}_{\text{ext}}=eE(t)\hat{x}$, where $e$ is the elementary charge. We use the time dependence
$eE(t)\equiv F\cos(\Omega(t-t_0))$, which is unbiased.

The system is also coupled to a thermal bath. We assume the transverse confinement to be strong enough so that the
probabilities of the direct bath-excited transitions between the transverse modes are negligibly small. Thus the
environment couples to the electronic degrees of freedom only through $\hat{x}$. Furthermore, in the spirit of the
Caldeira and Leggett model \cite{Caldeira}, we consider a harmonic bath with bilinear system-bath coupling.

The dynamical quantities of interest are the ratchet charge and spin currents $J_\text{C,S}(t)$ given as the statistical
average of the longitudinal charge and spin current operators,
$J_\text{C,S}(t)\equiv\text{Tr}[\hat{J}_\text{C,S}\hat{\rho}(t)]$, where $\hat{\rho}(t)$ is the reduced statistical
operator of the system, that is the full one with the bath degrees of freedom traced out. The CC operator is
$\hat{J}_\text{C}(t)=-ed\hat{x}/dt$ and for the SC operator we use the definition suggested in Ref. \cite{Shi},
$\hat{J}_\text{S}(t)=d\bigl(\hat{\sigma}_z\hat{x}\bigl)/dt$.

It is convenient to calculate the traces using the basis which diagonalizes both $\hat{x}$ and $\hat{\sigma}_z$, because
this requires to determine only the diagonal elements of the reduced density matrix. As shown in Ref. \cite{Smirnov}, for
a periodic system with RSOI the energy spectrum can be derived from the corresponding truly 1D problem without RSOI. This
leads to so-called Bloch sub-bands. The 2DEG is assumed to be sufficiently dilute to neglect the Pauli exclusion principle
in the temperature range of our problem. The upper limit of this temperature range is considered to be low enough so that
only the lowest Bloch sub-bands are populated. The basis which diagonalizes $\hat{x}$ and $\hat{\sigma}_z$ becomes in this
case discrete. The total number of the Bloch sub-bands is equal to the product of the number, $N_\text{B}$, of the lowest
Bloch bands from the corresponding truly 1D problem without RSOI, the number, $N_\text{t}$, of the lowest transverse modes
and the number of spin states. In this work we shall use the model with $N_\text{B}=1$, $N_\text{t}=2$. The total number of
the Bloch sub-bands in our problem is thus equal to four. Using $N_\text{B}=1$ we also assume that the external field is
weak enough and does not excite electrons to higher Bloch bands. The representation in terms of the eigen-states of
$\hat{x}$ for a model with discrete $x$-values is called discrete variable representation (DVR) \cite{Grifoni,Harris}. Let
us call $\sigma$-DVR the representation in which both the coordinate and spin operators are diagonal. Denoting the
$\sigma$-DVR basis states as $\{|\alpha\rangle\}$ and eigen-values of $\hat{x}$ and $\hat{\sigma}_z$ in a state
$|\alpha\rangle$ by $x_\alpha$ and $\sigma_\alpha$, respectively, the CC and SC are rewritten as
$J_\text{C}(t)=-e\sum_\alpha x_\alpha\dot{P}_\alpha(t)$ and $J_\text{S}(t)=\sum_\alpha\sigma_\alpha x_\alpha\dot{P}_\alpha(t)$,
where $P_\alpha(t)\equiv\langle\alpha|\hat{\rho}(t)|\alpha\rangle$ is the population of the $\sigma$-DVR state
$|\alpha\rangle$ at time $t$.

We are interested in the long time limit $\bar{J}^\infty_\text{C,S}$ of the currents $\bar{J}_\text{C,S}(t)$, averaged over
the driving period $2\pi/\Omega$.

The advantage of working in the $\sigma$-DVR basis is that real-time path integral techniques can be used to trace out
exactly the bath degrees of freedom \cite{Grifoni_1,Weiss}. Moreover, at driving frequencies larger than the ones
characterizing the internal dynamics of the quasi-1D system coupled to the bath, the averaged populations
$\bar{P}_\alpha(t)$ can be found from the master equation,
\begin{equation}
\dot{\bar{P}}_\alpha(t)=\sum_{\beta,(\beta\neq\alpha)}\bar{\Gamma}_{\alpha\beta}\bar{P}_\beta(t)-
\sum_{\beta,(\beta\neq\alpha)}\bar{\Gamma}_{\beta\alpha}\bar{P}_\alpha(t),
\label{averaged_master_equation}
\end{equation}
valid at long times. In Eq. (\ref{averaged_master_equation}) $\bar{\Gamma}_{\alpha\beta}$ is an averaged transition rate
from the state $|\beta\rangle$ to the state $|\alpha\rangle$.

The first task is thus to identify the $\sigma$-DVR basis. The eigen-states $|l,k_\text{B},j,\sigma\rangle$ of
$\hat{\sigma}_z$ were found in \cite{Smirnov} for the case $\gamma=0$. The results obtained in \cite{Smirnov} are
straightforwardly generalized to our model since for $N_\text{t}=2$ the operator $\hat{z}^2$ (and any even power of
$\hat{z}$) is effectively diagonal. The quantum numbers $l$, $k_\text{B}$, $j$, $\sigma$ stand for the Bloch band index,
quasi-momentum, transverse mode index and $z$-projection of the spin, respectively. As mentioned above $l=1$, $j=0,1$. One
further finds
\begin{equation}
\begin{split}
&\langle l',k_\text{B}',j',\sigma'|\hat{x}|l,k_\text{B},j,\sigma\rangle=\\
&=\delta_{j',j}\delta_{\sigma',\sigma}\;\;
{_j}\langle l',k_\text{B}'+\sigma k_\text{so}|\hat{x}|l,k_\text{B}+\sigma k_\text{so}\rangle_j,
\end{split}
\label{x_l_kb_j_sigma}
\end{equation}
where the index $j$ under the bra- and ket-symbols indicates that the corresponding electronic states are obtained using
the periodic potential $U_{\gamma,j}(x)\equiv U(x)[1+\gamma\hbar(j+1/2)/m\omega_0 L^2]$. For a fixed value of $j$ the
diagonal blocks in Eq. (\ref{x_l_kb_j_sigma}) are unitary equivalent and thus the eigen-values of $\hat{x}$ do not depend
on $\sigma$. The eigen-values of the matrix ${_j}\langle l',k_\text{B}'|\hat{x}|l,k_\text{B}\rangle_j$ are analytically
found and have the form $x_{\zeta,m,j}=mL+d_{\zeta,j}$, where $m=0,\pm1,\pm2\ldots$, $\zeta=1,2,\ldots,N_\text{B}$ and the
eigen-values $d_{\zeta,j}$ are distributed within one elementary cell. Thus one can label the eigen-states of $\hat{x}$ as
$|\zeta,m,j,\sigma\rangle$. The corresponding eigen-values are $x_{\zeta,m,j,\sigma}=x_{\zeta,m,j}$. We see that the
$\sigma$-DVR basis states $|\alpha\rangle$ introduced above are just the $|\zeta,m,j,\sigma\rangle$ states, that is
$\{|\alpha\rangle\}\equiv\{|\zeta,m,j,\sigma\rangle\}$.

To calculate CC and SC we use the tight-binding approximation assuming that the matrix elements
$\langle\zeta',m',j',\sigma'|\hat{H}|\zeta,m,j,\sigma\rangle$ with $|m'-m|>1$ are negligibly small. Let us introduce the
definitions for the states $|m,\xi\rangle\equiv|\zeta=1,m,\xi\rangle$ where $\{\xi\}=\{(j,\sigma)\}$ and
$\xi=1\Leftrightarrow(0,1)$, $\xi=2\Leftrightarrow(0,-1)$, $\xi=3\Leftrightarrow(1,1)$, $\xi=4\Leftrightarrow(1,-1)$.
Correspondingly, we introduce hopping matrix elements $\Delta_{\xi',\xi}^{m',m}\equiv\langle m',\xi'|\hat{H}|m,\xi\rangle$
($m'\neq m$ and/or $\xi'\neq\xi$) and on-site energies $\varepsilon_\xi\equiv\langle m,\xi|\hat{H}|m,\xi\rangle$.

Due to the harmonic confinement and RSOI the system is split into two channels: one with $\xi=1,4$ and another with
$\xi=2,3$. The two channels are independent of each other, that is, transitions between them are forbidden. This picture
is general and valid for an arbitrary number of the transverse modes. For clarity we below only consider the channel with
$\xi=1,4$. Two independent channels were also found for a different type of confinement in Ref. \cite{Perroni}.

Assuming that the hopping matrix elements are small enough we can use the second-order approximation \cite{Grifoni} for
the averaged transition rates in Eq. (\ref{averaged_master_equation}). We have
\begin{equation}
\begin{split}
&\bar{\Gamma}_{\xi'\!,\xi}^{m'\!,m}\!\!\!\!=\!\!
\frac{|\Delta_{\xi'\!,\xi}^{m'\!,m}|^2}{\hbar^2}
\!\!\!\!\int_{-\infty}^{\infty}\!\!\!\!\!\!\!\!\!d\tau\!\exp\!\biggl[\!-\frac{(x_{m,\xi}\!-\!x_{m',\xi'})^2}{\hbar}
Q[\tau,\!J(\omega)]+\\
&+\text{i}\frac{\varepsilon_\xi-\varepsilon_{\xi'}}{\hbar}\tau\biggl]
J_0\biggl[\frac{2F(x_{m,\xi}-x_{m',\xi'})}{\hbar\Omega}\sin\biggl(\frac{\Omega\tau}{2}\biggl)\biggl],
\end{split}
\label{transition_rate}
\end{equation}
where $x_{m,\xi}\equiv x_{\zeta=1,m,\xi}=mL+d_\xi$ with $d_\xi\equiv d_{1,j}$. In Eq. (\ref{transition_rate}) $J_0(x)$ denotes
the zero-order Bessel function and $Q[\tau,J(\omega)]$ is the twice integrated bath correlation function being a function
of time $\tau$ and a functional of the bath spectral density $J(\omega)$ \cite{Grifoni,Weiss}. The dependence of the
transition rates on the orbit-orbit coupling $\gamma$ comes from two sources. The first one is the Bloch amplitudes and
the second is the difference $\Delta d\equiv d_{1,0}-d_{1,1}$. In a tight-binding model the periodic potential is strong
and thus $\Delta d$ can be made less than all the relevant length scales, $\Delta d/l_\text{r}\ll 1$, where
$l_\text{r}=\text{min}[L,\:\sqrt{\hbar/m\omega_0},\:\hbar\Omega/F]$. Hence the main effect of the orbit-orbit coupling on
$\bar{\Gamma}_{\xi',\xi}^{m',m}$ comes only through the Bloch amplitudes, and we neglect terms of order
$\mathcal{O}(\Delta d/l_\text{r})$.

We then arrive at the main results of our work, the absence of the charge transport, $\bar{J}_\text{C}^\infty=0$, and the
expression for the non-equilibrium spin current (NESC),
$\bar{J}^\infty_\text{n-e,S}\equiv \bar{J}_\text{S}^\infty-\bar{J}_\text{e,S}^\infty$:
\begin{equation}
\begin{split}
\bar{J}^\infty_\text{n-e,S}=&-2L\biggl(\frac{I_{14}I_{41}}{I_{14}+I_{41}}-
\frac{I^{(0)}_{14}I^{(0)}_{41}}{I^{(0)}_{14}+I^{(0)}_{41}}\biggl)\frac{k_\text{so}^2\hbar^3\omega_0}{m}\times\\
&\times\sum_{k_\text{B},k_\text{B}'}\sin[(k_\text{B}-k_\text{B}')L]\text{Im}[\mathcal{F}_{k_\text{B},k_\text{B}'}],
\end{split}
\label{stationary_averaged_spin_current_b}
\end{equation}
where $I_{\xi',\xi}$,  $I^{(0)}_{\xi',\xi}$ are the integrals from (\ref{transition_rate}) with and without driving,
$F\neq 0$ and $F=0$, respectively, and
\begin{equation}
\begin{split}
  \mathcal{F}_{k_\text{B},k_\text{B}'}&\equiv u_{\gamma,0;1,k_\text{B}+k_\text{so}}^\text{DVR}(d_{1,0})
  u_{\gamma,1;1,k_\text{B}'-k_\text{so}}^\text{DVR}(d_{1,1})\times\\
  &\times[u_{\gamma,1;1,k_\text{B}-k_\text{so}}^\text{DVR}(d_{1,1})u_{\gamma,0;1,k_\text{B}'+k_\text{so}}^\text{DVR}(d_{1,0})]^*,
\end{split}
\label{F_function}
\end{equation}
where $u_{\gamma,j;1,k_\text{B}}^\text{DVR}(d_{1,j})$ is the DVR Bloch amplitude of the first band for electrons in the
periodic potential $U_{\gamma,j}(x)$.

In Eq. (\ref{stationary_averaged_spin_current_b}) we have eliminated from $\bar{J}_\text{S}^\infty$ the equilibrium spin
current (ESC), $\bar{J}_\text{e,S}^\infty$, following Ref. \cite{Rashba_1}. The fact that the ESC turns out to be finite
shows that the definition of SC suggested in Ref. \cite{Shi} does not automatically eliminate the presence of ESC.
However, as pointed out in Ref. \cite{Shi}, this current really vanishes in insulators. This can be seen from Eq.
(\ref{stationary_averaged_spin_current_b}). When the potential is strong, electrons are localized, the dependence of
the function $\mathcal{F}_{k_\text{B},k_\text{B}'}$ on the quasi-momentum disappears, and as a result both ESC and NESC are
equal to zero. This reasonable result is ensured by the spin current definition taking proper care of the spin torque. It
is interesting to note that ESCs are present even in a system with strong dissipation. As recently proposed in Ref.
\cite{Sonin}, ESCs can effectively be measured using a Rashba medium deposited on a flexible substrate playing a role of a
mechanical cantilever.
\begin{figure}
\includegraphics[width=6.4 cm]{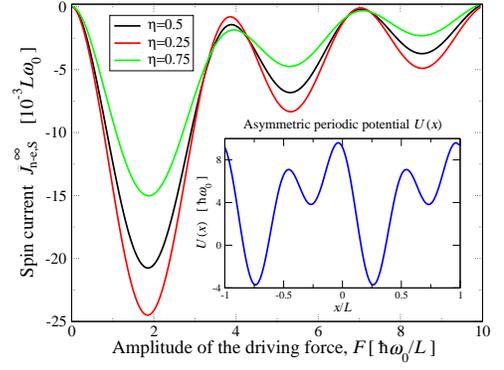}
\caption{\label{figure_2} (Color online) Non-equilibrium spin current, $\bar{J}^\infty_\text{n-e,S}$, as a function of the
amplitude, $F$, of the driving force for different values of the viscosity coefficient $\eta$. Temperature
$k_\text{Boltz.}T=0.5$, spin-orbit coupling strength $k_\text{so}L=\pi/2$, orbit-orbit coupling strength $\gamma=0.1$,
driving frequency $\Omega=1$. The inset displays the shape of the periodic potential.}
\end{figure}

We can determine the conditions under which the SC is finite. First of all from Eq.
(\ref{stationary_averaged_spin_current_b}) it follows that the spin-orbit coupling must be finite, {\it i.e.}
$k_\text{so}\neq 0$. Further, from Eq. (\ref{F_function}) one observes, that when $\gamma=0$, the Bloch amplitudes do not
depend on $j$, $u_{\gamma=0,j;1,k_\text{B}}^\text{DVR}(d_{1,j})\equiv u_{1,k_\text{B}}^\text{DVR}(d_{1})$, and since
$[u_{1,k_\text{B}}^\text{DVR}(d_{1})]^*=u_{1,-k_\text{B}}^\text{DVR}(d_{1})$ (time-reversal symmetry), the function
$\mathcal{F}_{k_\text{B},k_\text{B}'}$ becomes even with respect to its arguments. Then from Eq.
(\ref{stationary_averaged_spin_current_b}) one gets zero SC. Thus the second condition is the presence of the orbit-orbit
coupling. Finally, since for a symmetric periodic potential the Bloch amplitudes are real functions, we conclude that the
function $\mathcal{F}_{k_\text{B},k_\text{B}'}$ is also real in this case, that is
$\text{Im}[\mathcal{F}_{k_\text{B},k_\text{B}'}]=0$. As a result the third condition is the presence of spatial asymmetry.

Below we present corresponding numerical results. All energies and frequencies are given in units of $\hbar\omega_0$ and
$\omega_0$, respectively. The parameters are taken for an InGaAs/InP quantum wire: $\hbar\omega_0=0.9$ meV;
$\alpha\equiv\hbar^2k_\text{so}/m=9.94\cdot10^{-12}$ eV$\cdot$m; $m=0.037m_0$, respectively. For $k_\text{so}L=\pi/2$ one
gets $L=0.32$ $\mu\text{m}$.

The dependence of the NESC on the amplitude of the external driving is shown in Fig.~\ref{figure_2} for the asymmetric
periodic potential (see inset) $U(x)=\sum_{n=0}^2 V_n\cos(2\pi nx/L-\phi_n)$ with $V_0=4$, $V_1=-V_0$, $V_2=3.89$,
$\phi_0=\phi_2=0.0$, $\phi_1=1.9$. The gap between the Bloch bands with $l=1$ and $l=2$ is
$\Delta E_{12}\thickapprox 10.5$. In Fig.~\ref{figure_2} $FL,\,\hbar\Omega<\Delta E_{12}$ that is the numerical results are
consistent with the theoretical model assumptions. As an example we have used an Ohmic bath with the spectral density
$J(\omega)=\eta\omega\exp(-\omega/\omega_c)$, where the viscosity coefficient (in units of $m\omega_0$) is
$\eta=0.25,\;0.5,\;0.75$, and the cutoff frequency is $\omega_c=10$. As it can be seen, the NESC has an oscillating nature.
However, the oscillation amplitude goes down when the driving increases. Physically such behavior can be attributed to an
effective renormalization of the band structure in a high-frequency electric field \cite{Grifoni_1}. The group velocity
decreases in a non-monotonous way which due to RSOI slows down the spin kinetics. For increasing values of $\eta$ the
dissipation induced decoherence in the system gets more pronounced. The system becomes more classical and thus the
tunneling processes become less intensive. This leads to the spin current reduction which one observes in
Fig.~\ref{figure_2}.
\begin{figure}
\includegraphics[width=6.4 cm]{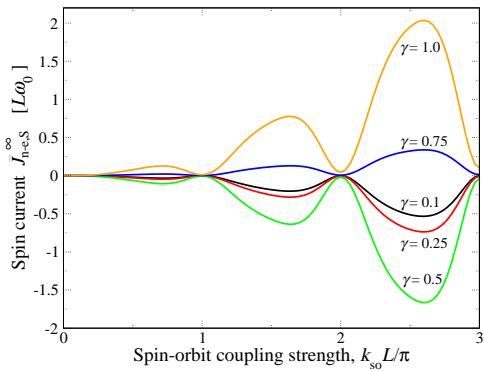}
\caption{\label{figure_3} (Color online) Non-equilibrium spin current, $\bar{J}^\infty_\text{n-e,S}$, as a function of the
spin-orbit coupling strength, $k_\text{so}$, for different values of the orbit-orbit coupling strength, $\gamma$. The
driving amplitude and viscosity coefficient are $F=2\hbar\omega_0/L$, $\eta=0.5$. The other parameters are as in
Fig.~\ref{figure_2}.}
\end{figure}

In Fig.~\ref{figure_3} the NESC is plotted versus $k_\text{so}L$ while $\gamma$ plays the role of a parameter. The
oscillations of the NESC have minima located at $nG/2$ where $n=0,1,2,\ldots$, and $G$ is the reciprocal lattice vector.
Physically this reflects the fact that for those values of $k_\text{so}$ the Rashba split becomes minimal due to the
periodicity of the energy spectrum in the $\vec{k}$-space. The magnitude of these oscillations decreases with decreasing
orbit-orbit coupling, and the current vanishes for $\gamma=0$.

In summary, we have studied stationary quantum transport in a driven dissipative periodic quasi-one-dimensional system
with Rashba spin-orbit interaction and orbit-orbit coupling. The spin ratchet effect has been investigated and an
analytical expression for the spin current has been derived and analyzed. This analysis has revealed that for the case of
moderate-to-strong dissipation the necessary conditions for non-vanishing spin currents are the spatial asymmetry of the
periodic potential as well as a finite strength of the spin-orbit interaction and orbit-orbit coupling. It has been
demonstrated that in a dissipative system equilibrium spin currents can exist. Our numerical calculations have shown
characteristic oscillations of the spin current as a function of the amplitude of the driving force and the spin-orbit
coupling strength. Finally, we note, that since the spin current has the in-plane polarization, it can be efficiently
measured by a magneto-optic Kerr microscope using the cleaved edge technology as suggested recently in Ref.
\cite{Kotissek}.

\begin{acknowledgments}
We thank J. Peguiron for useful discussions. Support from the DFG under the program SFB 689 is acknowledged.
\end{acknowledgments}

\end{document}